\documentclass[aps,pra,twocolumn,showpacs]{revtex4}
\usepackage{graphicx}
\usepackage{amsmath}
\newcommand{\er}{\mathbf{r}}
\newcommand{\de}{\mathrm{d}}
\newcommand{\ee}{{\rm e}}

\begin{document}

\title{Spin-Domain Formation in Antiferromagnetic Condensates}

\author{Michal Matuszewski}
\affiliation{Nonlinear Physics Center and ARC Center of Excellence
for Quantum-Atom Optics, Research School of Physical Sciences and
Engineering, Australian National University, Canberra ACT 0200,
Australia}

\author{Tristram J. Alexander}
\affiliation{Nonlinear Physics Center and ARC Center of Excellence
for Quantum-Atom Optics, Research School of Physical Sciences and
Engineering, Australian National University, Canberra ACT 0200,
Australia}

\author{Yuri S. Kivshar}
\affiliation{Nonlinear Physics Center and ARC Center of Excellence
for Quantum-Atom Optics, Research School of Physical Sciences and
Engineering, Australian National University, Canberra ACT 0200,
Australia}

\begin{abstract}
Antiferromagnetic condensates are generally believed not to display
modulational instability and subsequent spin-domain formation.  Here we demonstrate that in the
presence of a homogeneous magnetic field antiferromagnetic spin-1 Bose-Einstein
condensates can undergo spatial modulational instability followed
by the subsequent generation of spin domains. Employing numerical
simulations for realistic conditions, we show how this novel effect
can be observed in sodium condensates confined in an optical trap.
Finally, we link this instability and spin-domain formation with  stationary modes of the condensate.
\end{abstract}
\pacs{03.75.Lm, 05.45.Yv}

\maketitle

\section{Introduction}

The appearance of spin degrees of freedom in atomic matter waves
opens up possibilities for new phenomena such as spin
waves~\cite{Ho}, spontaneous magnetization~\cite{StamperKurn} and
spin mixing~\cite{Chapman}. However, perhaps the most intriguing
effect is associated with complex patterns, such as spin
textures~\cite{Ketterle2} or domains~\cite{Ketterle}, which may
appear either as stationary low-energy states or emerge
spontaneously due to condensate instabilities. Pattern
formation is a common feature in the dynamics of extended nonlinear
systems ranging from optics~\cite{book} to fluids~\cite{review}.
Such patterns often develop through the exponential growth of
unstable spatial modulations, known as {\em modulational
instability}.  In the spinor condensates we have the opportunity to
examine such effects in an environment which is remarkably easy to
control and manipulate, simply through the addition of an external
magnetic field.

The origin of the intriguing physics of spinor condensates lies in
the spin interaction between atoms, which allows for an exchange of
atoms between different spin components.  The parametric nature of
this interaction mirrors similar effects observed in nonlinear
optics, where the interaction of several optical modes may lead to
the development of new frequencies~\cite{book}.  Of particular
interest to our case is the possibility that instabilities of an
intense light beam may occur even when the wave is coupled to a
spatially stable eigenmode and propagates in the normal-dispersion
regime~\cite{wabnitz}; in this case the interplay of natural and
self-induced birefringence leads to nonlinear polarization symmetry
breaking and {\em polarization modulational instability}.  By
analogy we thus might expect similar instabilities in an initially
stable polar condensate subjected to additional spin component
coupling through an external magnetic field.

In the absence of an external magnetic field, the development of spatial modulational instability in three-component (or spin-1) {\em ferromagnetic} condensates and the subsequent formation of spin domains has been well established both theoretically~\cite{Robins,Saito2,DI} and experimentally~\cite{Quenched}.  However, early work on the zero field case~\cite{Robins} determined that antiferromagnetic (or polar) condensates are {\em modulationally stable}.  Experimental observations suggested this is also true for a weak magnetic field, however these experiments were carried out with a condensate smaller than a spin domain~\cite{black}.

In this paper we reveal that in fact the presence of a weak magnetic
field ($\sim 175\,$mG) leads to spin domain formation in antiferromagnetic condensates,
provided the condensate is larger than the spin healing length.  Furthermore we show that this spin domain formation is initiated by a {\em new type of modulational instability}, reminiscent of instabilities observed in nonlinear optics~\cite{wabnitz} and not seen before in Bose-Einstein condensates.  While spin-domain
formation in antiferromagnetic condensates has been observed before in the presence of a magnetic
field gradient~\cite{Ketterle}, we show here that it occurs equally well in the presence of a
homogeneous magnetic field.  Furthermore we reveal that this
modulational instability and spontaneous spin-domain formation is associated with stationary
states which exist in the presence of the weak magnetic field, and
which intrinsically break the validity of the single-mode
approximation (as seen earlier in \cite{NJP}).  We discuss realistic experimental 
conditions for the observation of these novel effects.

The paper is organized as follows. Section~\ref{sec_model} introduces theoretical model
of spin-1 condensate in a homogenous magnetic field. In Sec.~\ref{sec_mi} we investigate 
homogenous stationary states in magnetic field and analyze their stability with respect
to plane wave perturbations (modulational stability). Section~\ref{sec_dynamics}
presents results of numerical simulations corresponding to experimentally relevant
condensate evolution, demonstrating the possibility of observation of 
new instability in antiferromagnetic condesate. In Sec.~\ref{sec_states} we link this 
instability and spin-domain formation with  stationary modes of the condensate, and
Sec.~\ref{sec_conclusions} conludes the paper.

\section{Model} \label{sec_model}

The evolution of a dilute spin-1 ($F = 1$) Bose-Einstein condensate (BEC) in a
homogeneous magnetic field is given by the coupled Gross-Pitaevskii
equations,
\begin{align}
i \hbar\frac{\partial \Psi_{\pm}}{\partial t}&=\left[ \mathcal{L} +
\tilde{c}_2 (n_{\pm} + n_0 - n_{\mp})\right] \Psi_{\pm} +
\tilde{c}_2 \Psi_0^2 \Psi_{\mp}^* \,, \nonumber\\
i \hbar\frac{\partial \Psi_{0}}{\partial t}&=\left[ \mathcal{L} -
\delta E + \tilde{c}_2 (n_{+} + n_-)\right] \Psi_{0} + 2 \tilde{c}_2
\Psi_+ \Psi_- \Psi_{0}^* \,.
\label{GP}
\end{align}
where $\mathcal{L}= -\hbar^2\nabla^{2}/2m + \tilde{c}_0 n + V({\bf
r})$, $n_j=|\Psi_j|^2$, $n=n_+ + n_0 +n_-$, and $V({\bf r})$ is an
external potential.  The nonlinear coefficients are: $\tilde{c}_0=4
\pi \hbar^2(2 a_2 + a_0)/3m$ and $\tilde{c}_2=4 \pi \hbar^2(a_2 -
a_0)/3m$.  The total number of atoms $N=\int |n(\er)|^2 \de \er$ and
the total magnetization $M= \int \left[|n_+(\er)|^2 -
|n_-(\er)|^2\right]\de \er$ are conserved quantities. The Zeeman-energy shifts for each component can be calculated using the
Breit-Rabi formula~\cite{Wuster}
\begin{align}
E_{\pm}& = -\frac{1}{8}E_{\rm HFS}\left(1 + 4\sqrt{1\pm \alpha + \alpha^2} \right)  \mp g_I \mu_B B\,, \nonumber \\
E_{0} &= -\frac{1}{8}E_{\rm HFS}\left(1 + 4\sqrt{1 + \alpha^2} \right)\,,
\label{BR}
\end{align}
where $E_{\rm HFS}$ is the hyperfine energy splitting at zero
magnetic field, $\alpha = (g_I + g_J) \mu_B B/E_{\rm HFS}$, where
$\mu_B$ is the Bohr magneton, $g_I$ and $g_J$ are the gyromagnetic
ratios of electron and nucleus.  The linear part of the Zeeman
effect does not affect the condensate evolution, except for a change
in the relative phases~\cite{Beata} and so we remove it with the
transformation $\Psi_\pm \rightarrow \Psi_\pm\rm{exp}(-iE_\pm t)$,
$\Psi_0 \rightarrow \Psi_0\rm{exp}[-i(E_+ + E_-)t/2]$.  We thus
consider only the effects of the quadratic Zeeman shift, $\delta
E=(E_+ + E_- - 2E_0)/2  \approx \alpha^2 E_{\rm HFS}/16$, which is always
positive.

\section{Modulational instability of homogenous stationary states} \label{sec_mi}

First, we are interested in the stability analysis of the
homogeneous condensate and consider the case of vanishing potential,
$V({\bf r}) = 0$.  We look for the homogeneous solutions in the form
$\psi_{j} = \sqrt{n_{j}} \ee^{i\mu_{j}t + i \theta_j}$. The ``phase
matching condition" for Eqs.~(\ref{GP}) gives $\mu_+ + \mu_- = 2
\mu_0$. We find that both in the case of $B=0$ and in the case of
$M=0$, the steady state fulfills the stronger condition $\mu_+ =
\mu_- = \mu_0$.  However, if both magnetic field and magnetization
are nonzero, which is the case in real experiments, the chemical
potentials will be different, satisfying the less stringent phase
matching condition.

\begin{figure}[tbp]
\includegraphics[width=8cm]{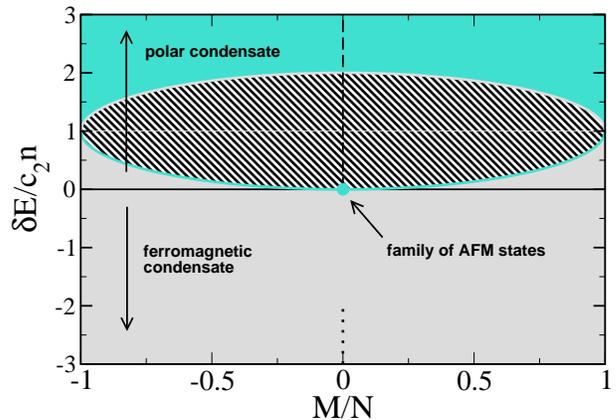}
\caption{(Color online) Diagram of existence of stationary states in
spin-1 condensates. In addition to the three-component phase-matched and
anti-phase-matched solutions shown on the diagram, two-component solutions with
$\rho_0=0$ exist for arbitrary $M$, and one-component solutions
$\rho_j=1$ exist with $j=-,0,+$.  Dark-shading (top) -- anti-phase-matched state,
light-shading (bottom) -- phase-matched state, cross-hatching --
both states. The dotted and dashed lines at $M=0$ indicate the
absence of a phase-matched or anti-phase-matched state respectively.
}\label{phase_diag}
\end{figure}

We define the density fraction in each component as $\rho_j=n_j/n$.
If we assume that all three spin components $\rho_j$ are
nonvanishing, the relative phase between them, $\theta = 2 \theta_0
- \theta_+ - \theta_-$, can take the value $0$ or $\pi$ only. We
will describe the corresponding stationary states as phase-matched ($\theta=0$)
and anti-phase-matched ($\theta=\pi$). Note that both types of states can in general exist in
both ferromagnetic and polar condensates \cite{Beata,Passos,GreeceDW}. 
However, phase-matched states are energetically favorable in ferromagnetic condensates,
and anti-phase-matched states in polar condensates \cite{Beata,NJP}. For that reason,
they were named ferromagnetic and polar states respectively in Ref.~\cite{Beata}. In
Fig.~\ref{phase_diag} we present the existence diagram for
three-component homogeneous stationary states.  For generality we also include the results for ferromagnetic condensates.  The ferromagnetic
condensates, such as $^{87}$Rb, occur in the lower half (where $c_2$
is negative), while polar condensates, such as $^{23}$Na, occur in
the upper half ($c_2$ is positive). There is clearly a region of
coexistence of anti-phase-matched and phase-matched states for a polar condensate
in nonzero magnetic field. In addition, a two-component solution with
$\rho_0=0$, and one-component solutions with $\rho_j=1$ exist. Our
results are in agreement with the previous analysis of homogeneous
ground states \cite{NJP}.

The energy density is related to the Hamiltonian of the system, from
which Eqs.~(\ref{GP}) are derived, by $H = \int E \de \er$. In
addition to the anti-phase-matched ground state \cite{NJP}, the polar condensate in the
coexistence region of Fig.~\ref{phase_diag} has an excited
phase-matched state corresponding to the energy maximum at
$\theta=0$. This state is stable with respect to spatially
homogeneous spin mixing, because the possible dynamical trajectories
in the $(\rho_0,\theta)$ plane correspond to a constant energy
value, hence both minima and maxima are stable.

The stability properties of these states
change when we consider the possibility of a spatial, or
modulational, instability (MI).  We calculate the growth rate of the
Bogoliubov modes \cite{Smerzi},
\begin{equation}
\psi_j=\left[\sqrt{n_j} + u_j(t) \ee^{i{\bf k \cdot x}} + v_j^*(t)
\ee^{-i{\bf k \cdot x}} \right] \ee^{i \mu_j t + i \theta_j}\,.
\end{equation}
After substituting the above to Eq.~(\ref{GP}) we obtain a set of
equations for the vector ${\bf z} = (u_+,u_0,u_-,v^*_+,v^*_0,v^*_-)$
in the form $\de {\bf z}/{\de t} = i A {\bf z}$, where $A$ is a
6th-rank matrix \cite{Trillo}. For an equilibrium state, it is
possible to eliminate $\mu_j$ and $\delta E$ from $A$, expressing it
in terms of $n_j$, $c_0$, $c_2$ and $k$ only for specific $\theta$.
The Bogoliubov modes are the solutions of the characteristic
equation $\det (A - \hbar\omega) = 0$, with $\omega$ being the
eigenfrequency of the excitation. The form of this equation is too
cumbersome to present here, in contrast to the $B=0$ case considered
before \cite{DI}. In general, one has to use numerical methods to
obtain a set of solutions.  

The general numerical results are
presented in a systematic way in Table~\ref{TMI}.  These results
have been calculated for the specific cases of the $^{87}$Rb
condensate (upper half) and $^{23}$Na condensate (lower half) with
scattering lengths given in \cite{Beata} and hyperfine energy
splitting given in \cite{Arimondo}.  As we are considering the
homogeneous case the result is applicable to one-, two- and
three-dimensional condensates.
\begin{table} [tbp]
\begin{tabular}{|l|l|c|c|c|c|}
\hline
condensate & state & $B=0$ & $B=0$ & $B\neq 0$ & $B\neq 0$  \\
type & type & $M=0$ & $M\neq 0$ &  $M=0$ & $M\neq0$ \\
\hline
 & PM & stable & stable & stable & stable \\
ferro & APM & unstable$^1$ & X & X & X \\
 & $\rho_0=0$ & unstable & unstable & unstable & unstable \\
 & $\rho_0=1$ & unstable & X & unstable & X \\
\hline
 & PM & stable & stable & unstable & unstable \\
polar & APM & stable$^{1,2}$ & X & X & unstable \\
 & $\rho_0=0$ & stable & stable & unstable & unstable \\
 & $\rho_0=1$ & stable & X & stable & X \\
\hline
\end{tabular}
\caption{Stability of spin-1 condensate states in absence and
presence of the magnetic field: PM - phase-matched, APM - anti-phase-matched, X -
state does not exist. $^1$A family of stationary states. $^2$Neutral
stability with respect to spatially homogeneous spin
mixing.}\label{TMI}
\end{table}
We do not consider here the trivial case of $\rho_{\pm}=1$ states,
which have extremal value of magnetization $M=\pm N$.
All the cases presented in this table were double checked by direct
numerical integration of Eq~(\ref{1D}) (see Sec.~\ref{sec_dynamics}).
The results cover the area of Fig.~\ref{phase_diag} 
and do not apply to the case of very high magnetic field, when 
the quadratic Zeeman energy dominates.

We find that our results are in agreement with the existing data for
the vanishing magnetic field case. In Ref.~\cite{Robins} the authors
found that all the condensate equilibrium states are stable, with
the exception of one state in the ferromagnetic condensate, which
corresponds to ``ferro-APM'', $M=0$ state in Table~\ref{TMI}. In
Ref.~\cite{Beata}, it was also found that ground states of both
ferromagnetic and polar condensates (``ferro-PM'' and ``polar-APM'') are
modulationally stable.  The authors of Ref.~\cite{DI} were considering
mainly non-equilibrium (spin-mixing) states, but their general
conclusion was that all polar condensate states are dynamically
stable, and ferromagnetic are not. Here we report that in fact
specific ferromagnetic condensate states are stable, while polar
condensates may become unstable.  Furthermore, as discussed below, we see
that dynamic domain formation may occur for polar condensates, and
lead to convergence to new stationary states.

\begin{figure}[tbp]
\includegraphics[width=8.5cm]{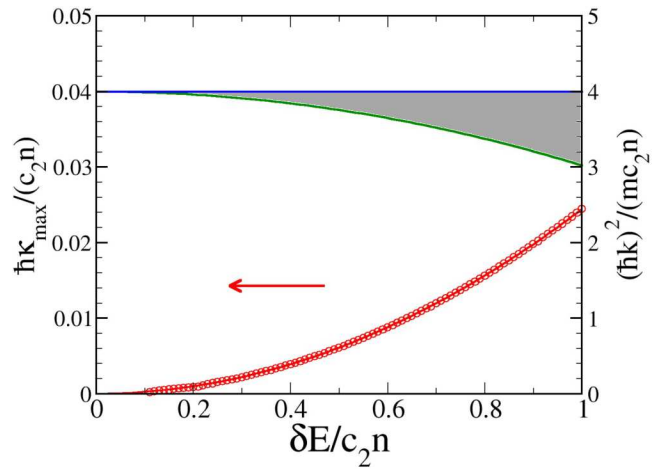}
\caption{(Color online) Modulational instability growth rate in a
phase-matched steady state of a sodium condensate (circles), versus
the quadratic Zeeman shift in the case $M=0$. The continuous line is
a square fit to the numerical data. The upper shaded area shows the
range of $k$ vectors corresponding to the unstable (imaginary)
frequencies.}\label{nagr}
\end{figure}

As one can see in Table~\ref{TMI}, the magnetic field affects
stability of polar condensates. We investigate this phenomenon in
detail by calculating the growth rate $\kappa={\rm Im}(\omega)$
corresponding to unstable Bogoliubov modes. In Fig.~\ref{nagr}, we
present results for a ``polar-PM'' state of a sodium condensate as a
function of the magnetic field strength. The growth rate is
proportional to the square of the quadratic Zeeman shift,
which is in turn proportional to the square of the magnetic field
strength. Hence, one has to apply a {\em relatively strong magnetic field}
to observe MI on a reasonable time scale.
Another interesting feature is the range of wavevectors $k$
corresponding to unstable modes. In contrast to the typical case
where this range starts from $k=0$, here the unstable region begins
at a nonzero minimum value. This type of ``optical mode'' branch has
been reported before in the case of parametric optical solitons
\cite{Trillo}.

\section{Dynamics of the condensate in a cigar-shaped trap} \label{sec_dynamics}

We now consider the implications of these results in the
experimentally relevant case of a $^{23}$Na condensate localized in
a cigar-shaped harmonic trap, $V({\bf r}) =
\frac{1}{2}m\omega_\perp^2 (y^2+z^2)+\frac{1}{2}m\omega_\parallel^2 x^2$.
Specifically we consider the case $\omega_\perp >> \omega_\parallel$
in which the Fermi radius of the transverse trapping potential is
smaller than the spin healing length, 
and nonlinear energy scale is much smaller than the transverse 
trap energy scale, which allows us to reduce the
problem to one spatial dimension \cite{Beata,NPSE}. Following
standard dimensionality reduction procedure we obtain the
one-dimensional model,
\begin{align}
\label{1D} i \hbar\frac{\partial \psi_{\pm}}{\partial t}&=\left[
\tilde{\mathcal{L}} + c_2 (n_{\pm} + n_0 - n_{\mp})\right]
\psi_{\pm} +
c_2 \psi_0^2 \psi_{\mp}^* \,, \nonumber\\
i \hbar\frac{\partial \psi_{0}}{\partial
t}&=\left[\tilde{\mathcal{L}} - \delta E + c_2 (n_{+} + n_-)\right]
\psi_{0} + 2 c_2 \psi_+ \psi_- \psi_{0}^* \,,
\end{align}
where $\tilde{\mathcal{L}} = -(\hbar^2/2m)\partial^{2}/\partial x^2 + c_0 n +
\frac{1}{2}m\omega_\parallel^2 x^2$ and the interaction coefficients
have been rescaled and now include the transverse trap frequency,
$c_0=4 \hbar\omega_{\perp}(2 a_2 + a_0)/3$ and $c_2=4
\hbar\omega_{\perp}(a_2 - a_0)/3$.  

\begin{figure}[tbp]
\includegraphics[width=8.5cm]{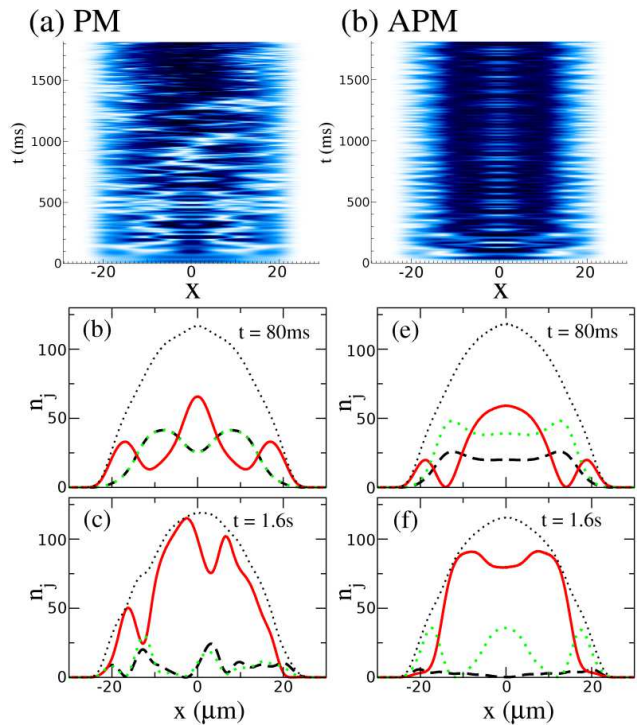}
\caption{(Color online) Spin-domain formation in a $^{23}$Na condensate
confined in an optical harmonic trap. Top panels: Evolution of $n_0$ (darker shading higher density). Bottom panels: densities at given times.
(a,b,c) phase-matched initial state, $\rho_{+,0,-} =
0.351,0.3,0.349$; (b) anti-phase-matched initial state, $\rho_{+,0,-} =
0.4,0.01,0.59$.  The thin dotted line corresponds to the total
condensate density $n$, and the dashed, solid and dotted lines
correspond to $n_+$, $n_0$ and $n_-$ respectively. Parameters
are $N=3.7\times 10^3$, $B=175\,$mG, $\omega_{\perp}= 2\pi\times
10^3\,$Hz, $\omega_{\parallel}= 2\pi\times 32\,$Hz.}\label{evolution}
\end{figure}

The experimental scenario we consider here consists of several phases.
Initially, the condensate is prepared in the $m=-1$ ground state
\cite{Quenched}. Next, short microwave field pulses are applied to
transfer the atomic population to the desired state \cite{Transfer}.
We consider two cases, a phase-matched state ($\theta = 0$) with
$\rho_{+,0,-} = 0.351,0.3,0.349$, and a anti-phase-matched state ($\theta=\pi$)
with $\rho_{+,0,-} = 0.4,0.01,0.59$.  Simultaneously,
the magnetic field is set to the value of $175\,$mG
\cite{Quenched}. The results of the corresponding numerical
simulations of Eqs.~(\ref{1D}) are presented in
Fig.~\ref{evolution}. The MI develops after
tens of milliseconds, and leads rapidly to spin-domain formation.
We see what appears to be initial
oscillations, followed by instability dynamics leading to an
apparent oscillating state.  In the case of the
phase-matched initial condition and almost zero magnetization we
ultimately see conversion to the $\psi_0$ component, as observed in experiment~\cite{black}.  In the anti-phase-matched
initial condition however, with nonzero magnetization, the spin
domains become well-defined and persist in the dynamics. We have 
verified that the nonlinear energy at peak density $c_0 n_{\rm max} / 2$ 
is much smaller than the transverse energy separation $\hbar \omega_{\perp}$,
which justifies the use of reduced 1D GPE \cite{NPSE}.

\section{Stationary spinor states} \label{sec_states}

\begin{figure}[tbp]
\includegraphics[width=8.5cm]{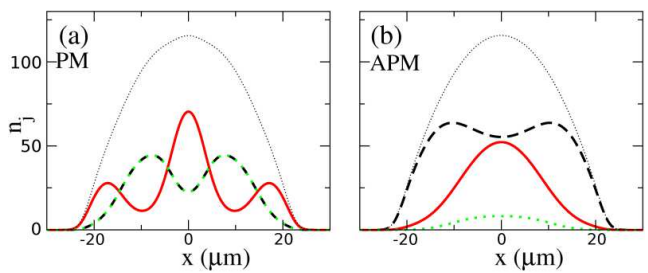}
\caption{(Color online) Examples of (a) phase-matched and (b) anti-phase-matched stationary states in a $^{23}$Na
condensate.  Shown are the total density $n$ (thin dotted line) and
component densities $n_{+,0,-}$  (dashed, solid and dotted lines respectively).}
\label{stationary}
\end{figure}

The existence of MI suggests that domain-type
stationary states may be expected in the trap, just as plane wave
instability and solitons are typically found together in optics.
Indeed, as can be seen in Fig.~\ref{stationary}, we find that in the
stationary picture the profiles always break the single-mode
approximation for both the phase-matched and anti-phase-matched states (as found
in Ref.~\cite{NJP} for the case of a anti-phase-matched state in a polar condensate).
As predicted by the MI analysis, an initially smooth profile will
therefore become modulated with the ensuing instability dynamics reflecting the nearby stationary state profiles.  For instance
comparing Fig.~\ref{evolution}(b) and Fig.~\ref{stationary}(a) we
see that profiles similar to the stationary states appear in the
evolution.  Significantly, but unsurprisingly in a polar condensate,
we find that the phase-matched state is generally unstable, while
the anti-phase-matched state is stable.  The stability of these states appears to
reflect the ultimate dynamics of the condensate with the
phase-matched state breaking up while the anti-phase-matched state has
highly persistent spin domains.  A complete analysis of the
stability of the stationary states will be presented elsewhere.

\section{Conclusions} \label{sec_conclusions}

We have demonstrated that an antiferromagnetic
spin-1 condensate can undergo a novel type of spatial modulational instability
followed by subsequent spin-domain formation in the presence of a homogeneous magnetic field. We have employed
realistic conditions to demonstrate, with the help of numerical
simulations, that this novel modulational instability can be
observed in a sodium condensate confined in an optical trap
potential and that the ensuing instability dynamics connect with the stationary states in the trap.

\acknowledgments

This work was supported by the Australian Research Council through
the ARC Discovery Project and Centre of Excellence for Quantum-Atom Optics.
M.~M.~acknowledges support from the Foundation for Polish Science.

\clearpage

\end{document}